# JULOC：A Local 3-D High-Resolution Crustal Model in South China for Forecasting Geoneutrino Measurements at JUNO

Ruohan Gao[1, *], Zhiwei Li[2, *], Ran Han[3, *], Andong Wang[4], Yufeng Li[5], Yufei Xi[6], Jingao Liu[1], Xin Mao[3], Yao Sun[1], Ya Xu[7,8]

1. *China University of Geosciences (Beijing), Beijing, 100083, China.*
2. *State Key Laboratory of Geodesy and Earth's Dynamics, Institute of Geodesy and Geophysics, Chinese Academy of Sciences, Wuhan, 430077, China.*
3. *Science and Technology on Reliability and Environmental Engineering Laboratory, Beijing Institute of Spacecraft Environment Engineering, Beijing 100094, China.*
4. *State Key Laboratory of Nuclear Resources and Environment, East China University of Technology, Nanchang, 330013, China.*
5. *Institute of High Energy Physics, Chinese Academy of Sciences, Beijing, 100049, China.*
6. *Institute of Hydrogeology and Environmental Geology, Chinese Academy of Geological Sciences, Shijiazhuang, 050061, China.*
7. *Key Laboratory of Petroleum Resource Research, Institute of Geology and Geophysics, Chinese Academy of Sciences, Beijing, 100029, China.*
8. *Institute of Earth Science, Chinese Academy of Sciences, Beijing, 100029, China.*

Corresponding Authors: Ruohan Gao (ruohangao@cugb.edu.cn)

Zhiwei Li (zwli@whigg.ac.cn)

Ran Han (hanran@ncepu.edu.cn)

**Key Points**

- Geoneutrino measurement at JUNO has the potential to discriminate different Bulk Silicate Earth models.
- We build a refined 3-D model of the local crust around JUNO combining geological, geophysical and geochemical information.
- Our method has the potential to significantly reduce the error of crustal geoneutrino signal estimation.

## Abstract

Geothermal energy is one of the keys for understanding the mechanisms driving the plate tectonics and mantle dynamics. The surface heat flux, as measured in boreholes, provides limited insights into the relative contributions of primordial versus radiogenic sources of the interior heat budget. Geoneutrino, electron antineutrino that is produced from the radioactive decay of the heat producing elements, is a unique probe to obtain direct information about the amount and distribution of heat producing elements in the crust and mantle. Cosmochemical, geochemical, and geodynamic compositional models of the Bulk Silicate Earth (BSE) individually predict different mantle neutrino fluxes, and therefore may be distinguished by the direct measurement of geoneutrinos. Due to low counting statistics, the results from geoneutrino measurements at several sites are inadequate to resolve the geoneutrino flux. However, the JUNO detector, currently under construction in South China, is expected to provide an exciting opportunity to obtain a highly reliable statistical measurement, which will produce sufficient data to address several vital questions of geological importance.

However, the detector cannot separate the mantle contribution from the crust contribution. To test different compositional models of the mantle, an accurate a-priori estimation of the crust geoneutrino flux based on a three-dimensional (3-D) crustal model is important. This paper presents a 3-D crustal model over a surface area of 10° × 10° grid surrounding the JUNO detector and a depth down to the Moho discontinuity, based on the geological, geophysical and geochemical properties. This model provides a distinction of the thickness of the different crustal layers together with the corresponding Th and U abundances. We also present our predicted local contribution to the total geoneutrino flux and the corresponding radiogenic heat. Compared to previous studies where the surface layer is subdivided into a few geologic units and each of them is considered to have the same geochemical property, our method has provided an effective approach to reduce the uncertainty of geoneutrino flux prediction by constructing the composition of the surface layer through cell by cell which are independent to each other.

**Contents**

## 1. Introduction

Radiogenic heat is an important ingredient for the Earth's interior dynamics and evolution, among which 99% comes from the radioactive decays of $^{238}U$, $^{232}Th$ and $^{40}K$. These three nuclides belong to three elements (i.e., U, Th and K) which are known as the Heat Producing Elements (HPEs). However, the estimated abundances of HPEs in the Bulk Silicate Earth (BSE), and thus the estimated amount of total radiogenic heat in the Earth, vary by about 3 times in different models (e.g., McDonough and Sun, 1995; Turcotte and Schubert, 2014; O'Neill and Palme, 2008; Arevalo et al., 2009; Javoy et al., 2010). The uncertainty mainly comes from the limited sampling access to the Earth's deep mantle.

Recent advances in experimental neutrino physics provide a new approach in probing deep Earth (Araki et al., 2005, Bellini et al., 2010, Donini et al., 2019). Geoneutrinos are produced by the beta decaying processes of radioactive nuclides, and travel through the Earth with nearly negligible interference (Mantovani et al., 2004; Fiorentini et al., 2007). The geoneutrino flux at the Earth's surface reflects the inventory and distribution of HPEs in the Earth's interior. Therefore, different mantle models would predict different surface geoneutrino fluxes, and may be discriminated by direct geoneutrino measurements at the Earth's surface. In 2005, the KamLAND collaboration published the first experimental result of geoneutrino measurement (Araki et al., 2005), and then Borexino collaboration claimed the second measurement (Bellini et al., 2010). However, the working geoneutrino detectors (KamLAND and Borexino) have not accumulated geoneutrino data with precision good enough to discriminate the different models that parameterize the composition of the BSE.

Now we have a new opportunity for a better geoneutrino measurement with the development of the Jiangmen Underground Neutrino Observatory (JUNO). The JUNO facilities will include a 20 kton liquid-scintillator detector (An et al., 2016), which is 20 times the size of the KamLAND detector and 60 times larger than the Borexino detector. JUNO will be able to detect a much larger number of geoneutrino events per year, and therefore can measure the Earth's total geoneutrino flux with higher precision (Han et al., 2016).

The directionality of geoneutrinos is presently not detectable by the liquid scintillator, which is only sensitive to the integrated flux of all directions. This means the experiment cannot separate the mantle contribution from the crust contribution. In order to enable a quantitative test of different compositional models of the mantle, an accurate estimation of the crust geoneutrino flux is required in advance (Dye, 2010; Šrámek et al., 2013). The geoneutrino flux from a source decreases with distance via the inverse square law. The regional crust within 500 km around the detector contributes more than 50% of the total geoneutrino signal (Mantovani et al., 2004). A refined 3-D model of the local crust is important for precise estimation of the crust geoneutrino flux. So far, the existing crustal model for JUNO geoneutrino estimation only focused on constructing a geophysical and gravimetric model of the local crust (Reguzzoni et al., 2019). Here we work on establishing a refined 3-D crustal model of the closest $10° \times 10°$ grid surrounding the detector, by integrating geological, geochemical, and geophysical data. This model is then used to calculate the crustal contribution of geoneutrino signal at the JUNO detector with uncertainty. We also discuss the way in building a crustal composition model cell by cell to improve the uncertainty of the estimated crust geoneutrino flux, so that the mantle geoneutrino flux (i.e., its HPEs abundances) can be better constrained with future geoneutrino measurements.

## 2. Research Area and Geologic Background

The JUNO detector is located on the coastline of South China (N 22.12°, E 112.52°). Our local 3-D crustal model focuses the closest 10° × 10° grid surrounding the detector (N 17-27°, E 107.5-117.5°). The study area is mainly comprised of the South China Block to the northwest and the South China Sea to the southeast (Fig. 1).

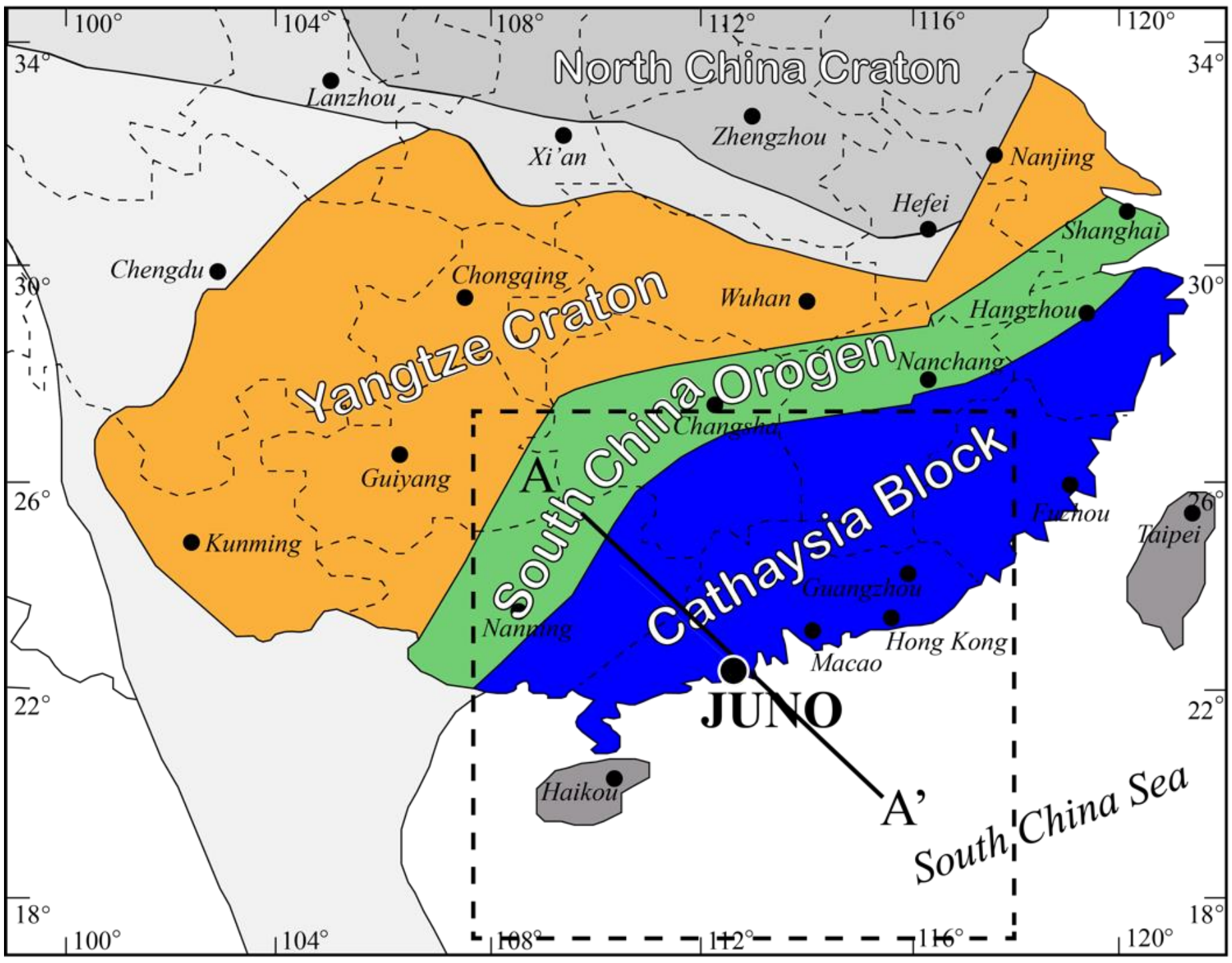

*Figure 1. Schematic map of the region around the JUNO detector. The research area of this study is marked by the dashed line rectangle. A-A' shows the position of the cross section in Fig. 3.*

The South China Block was constructed by two microplates: the Cathaysia and Yangtze Blocks. The Cathaysia Block is primarily composed of Paleoproterozoic basement and sporadic crustal fragments with Late Archean ages of ~2.5 Ga (e.g., Chen and Jahn, 1998). In contrast, the Yangtze Block has a stable Archean basement dated up to >3.2 Ga with an average age of 2.7-2.8 Ga (e.g., Gao et al., 1999). The Cathaysia and Yangtze Blocks collided and amalgamated during the Early Neoproterozoic (ca. 800-850 Ma), and the suture zone is also known as the Jiangnan Orogeny Belt (e.g., Li and Li, 2007).

The thickness of the South China continental crust varies between ~30-40 km (e.g., Hao et al., 2014). The continental crust is mainly composed of crystalline basement including meta-sedimentary and meta-igneous rocks, Precambrian-Quaternary sedimentary rocks with minor volcanic rocks, and granitic intrusions. The Precambrian-Quaternary sedimentary rocks can reach up to >5 km of thickness and include variable lithologies, such as sandstone, mudstone, carbonate, and minor volcanic rocks. Many of the sedimentary rocks have been affected by low grade metamorphisms (e.g., Gao et al., 1999). The granitic intrusions are mainly distributed along the coast, which are the results of three tectono-magmatic events. In the Early Paleozoic, intensive reworking of the South China crust has given rise to voluminous intraplate migmatites and granites (e.g., Faure et al., 2009). The collision between the South China Block and the North China Block in the Triassic also led to extensive magmatism (e.g., Chen and Jahn, 1998). Later on, large volumes of granites were emplaced in the Jurassic and Cretaceous periods as the result of regional compression or extension (e.g., Zhou et al., 2006). Overall, the ages of the granite intrusions increase towards inland.

The South China Sea can be divided into three parts, the northern continental margin, the central oceanic basin and the southern continental margin. Our research area covers most of the northern continental margin and a small portion of the oceanic basin. The oceanic basin is widely accepted to be formed by Paleogene continental rifting and subsequent oceanic floor spreading during the Oligo-Miocene (e.g., Holloway, 1982; Clift and Lin, 2001). The northern continental margin is a rifted continental margin that is formed by progressive extension of the lithosphere, which mostly resembles the style of a magma-poor margin but manifests voluminous late-stage volcanic rocks spanning a width of ~250 km along the continent-ocean transition zone (e.g., Clift and Lin, 2001; Pubellier et al., 2017).

The crust thickness of the South China Sea decreases southward, from about 30 km along the coast to 7-8 km in the central basin (e.g., Yan et al., 2001; Hao et al., 2014). The oceanic basin contains typical oceanic crust that consists of sediments and mafic igneous rocks. In comparison, the crust at the northern continental margin is more complicated. Sedimentation on the rifted continental crust basement started since Paleogene, with sediment thickness up to >5 km (e.g., Yan et al., 2001; Qiu et al., 2001). Mesozoic granitic intrusions are a common feature on the rifted continental margin, and are likely genetically related to the Mesozoic granites in the South China Block (e.g., Qiu et al., 2008; Pubellier et al., 2017).

## 3. 3-D crustal model from seismic ambient noise tomography in South China

We first attempt to construct a 3-D model of crust layering in South China via seismic ambient noise tomography. First, one-year continuous seismic ambient noise waveform data from 450 permanent, broadband seismic stations in South China and adjacent regions were collected from the China National Seismic Network (Data Management Centre of China National Seismic Network, 2007; Zheng et al., 2010) (Fig. A1 in Appendix A). The data processing of waveform cross-correlation calculations and quality control are done following the modified method of Bensen et al. (2007). One-hour waveform segments are used to remove the mean, trend and instrument response. Spectrum whitening and normalization are also conducted. In order to improve the Signal-to-Noise Ratios (SNRs), Noise Cross-correlation Functions (NCFs) are stacked for each station pair, then averaged by the positive and negative lags. Only those NCFs with SNR $\geq$ 8 are used in the dispersion curve measurement. The automatic frequency-time analysis method is used to measure the group and phase velocity dispersion curves of the Rayleigh waves (Shapiro et al., 2005; Bensen et al., 2007; Yang et al., 2011; Shen et al., 2016). After quality control, totally 41,041 dispersion curves for both group and phase velocities are used to conduct the surface wave tomography. An iterative non-linear tomographic method with fast-marching method and subspace inversion algorithm is used for surface wave tomography at periods of 6~40 s (Rawlinson and Sambridge, 2004) (Fig. A2 and A3 in Appendix A). After checkerboard resolution tests and spatial resolution analysis, resolutions at each period are estimated to be ~50 km or better for most areas of South China.

When constructing the 3-D crustal model in South China, the 3-D shear wave velocity model down to 60 km depth are inverted with the group and phase velocity results at all grids (Herrmann, 2013; Li et al., 2016) (Fig. A4 in Appendix A). Based on the inverted 3-D shear wave velocity model, we calculated the synthetic group and phase velocity maps to compare with the observed velocity maps. Good consistency is obtained for the synthetic and observed velocity maps, suggesting the 3-D shear wave velocity model can well fit the observations (Fig. A2 and A3 in Appendix A).

In order to obtain the layering (upper, middle and lower) of the crust, we calculated the P-wave velocity from the 3-D S-wave velocity model based on the empirical relations between P- and S-wave velocities proposed by Brocher (2005) (equation 9). After extensive tests and comparisons with that from the CRUST1.0 model (Laske et al., 2013), we chose isosurfaces of P-wave velocities of 6.0 km/s and 6.4 km/s, respectively, as the boundaries between the upper-mid crust and mid-lower crust in the areas with good resolutions of seismic ambient noise tomography in this study. Whereas in the regions where the resolutions are low for the seismic ambient noise tomography (e.g., the South China Sea area), the crustal layering of CRUST1.0 model is adopted. As for the Moho depth, the model derived from Bouguer gravity inversion under the constraints of ~120 deep seismic sounding profiles is preferred (Hao et al., 2014), which provides better resolution and coverage of the research area than other available models (e.g., CRUST1.0; He et al., 2014; Xia et al., 2015). Fig. A5 in Appendix A shows the crust layering of the area about 500 km spanning from JUNO.

Moreover, the empirical relations between P-wave velocity and density are used to build the 3-D density model (Brocher, 2005) (equation 3). Fig. 3 shows the cross section traversing the JUNO site in NW-SE direction (Line A-A' in Fig. 1), including information of shear-wave velocity and density structures with uncertainties. The depth boundaries of the upper-mid and mid-lower crust are ~15 and ~ 20 km beneath JUNO, respectively. The crustal thickness is ~ 30 km, which is close to the average crustal thickness of the South China Block. The crustal thickness gradually decreases from ~40 km in the southeast margin of Tibet Plateau to ~15 km in the north part of the South China Sea.

In addition, bootstrap method is used to examine the errors in the tomographic inversion (Koch, 1992; Hearn and Ni, 1994) and then evaluate the uncertainty of the 3-D crustal model. All group and phase velocity dispersions are randomly selected with uniform distribution, and keep the total numbers of dispersion curves the same as that used in the real inversion. With 100 trial inversions, the standard derivations of shear wave velocities are calculated and considered as the uncertainty of the geophysical model. For most areas in South China, the uncertainties in shear wave velocity are mostly less than 0.03 km/s at depths from 5 ~ 45 km (Fig. A6 in Appendix A), and the estimated uncertainties for the density structures are less than 0.04 g/cm$^3$ for most area in the crust. Other factors, such as the relation used to convert P-wave velocities to densities, are very difficult to consider quantitatively in the density uncertainty estimation. However, for the crustal structure, the stratified-homogeneous characteristics are well constrained by seismic data. For example, as shown in Christense & Mooney (1995, JGR), the global average crustal P-wave velocity is 6.45 km/s ± 0.21 km/s, with about ~ 3.3% uncertainty, suggesting a similar density uncertainty. As for the South China block, the crustal structure is pretty uniform, no significant large-scale velocity and density anomalies are found in previous studies, and thus the density uncertainty should be less than that of global crustal model.

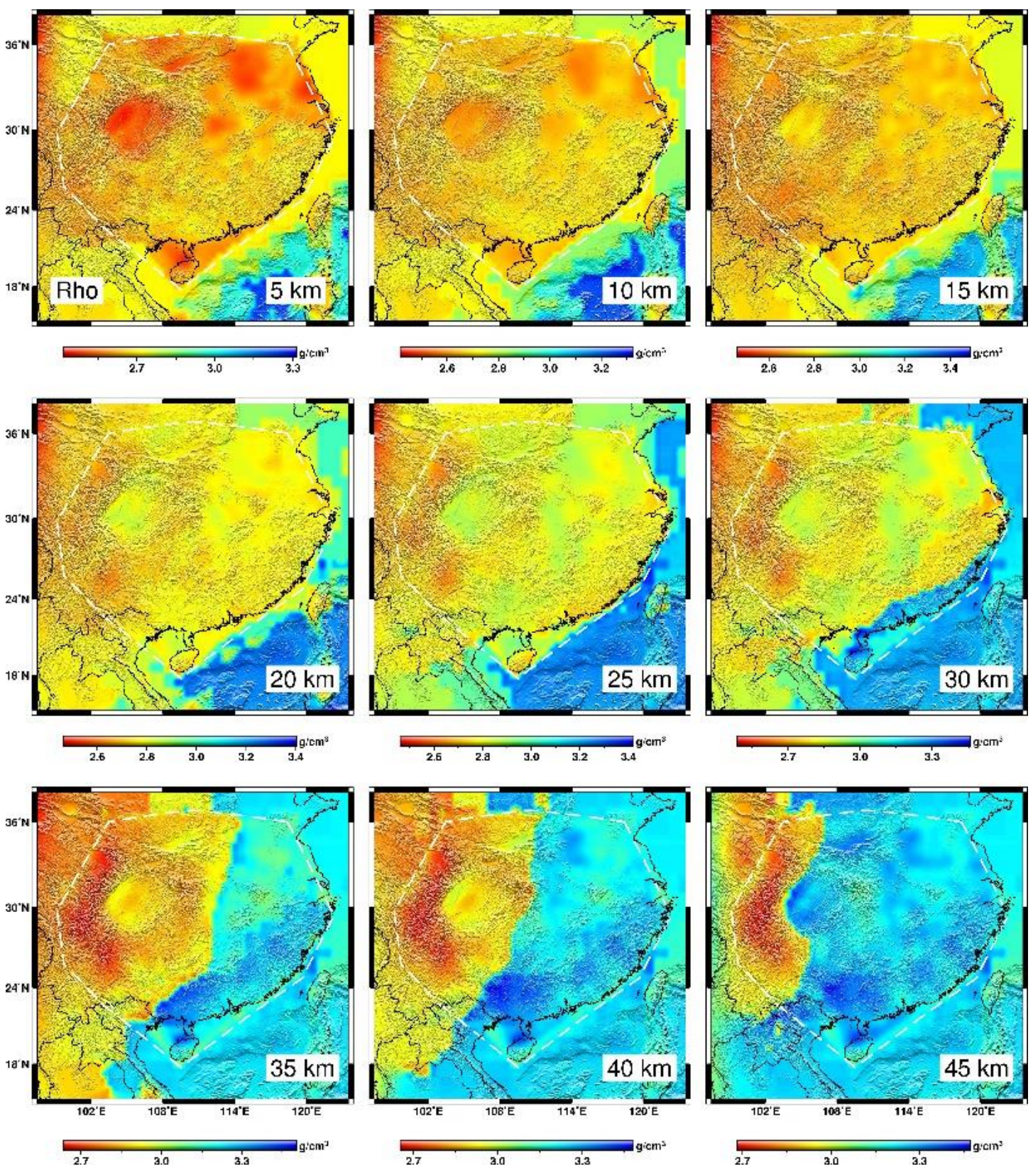

***Figure 2.*** *3-D Density structures in the South China Block and surrounding regions. The density structures are derived from the 3-D velocity model of seismic ambient noise tomography based on the empirical relations between the density and velocity for crustal materials (Brocher, 2005). The white dashed line shows the regions with high-resolution results derived from seismic ambient noise tomography. For the outside regions, the structures are derived from the global CRUST1.0 model.*

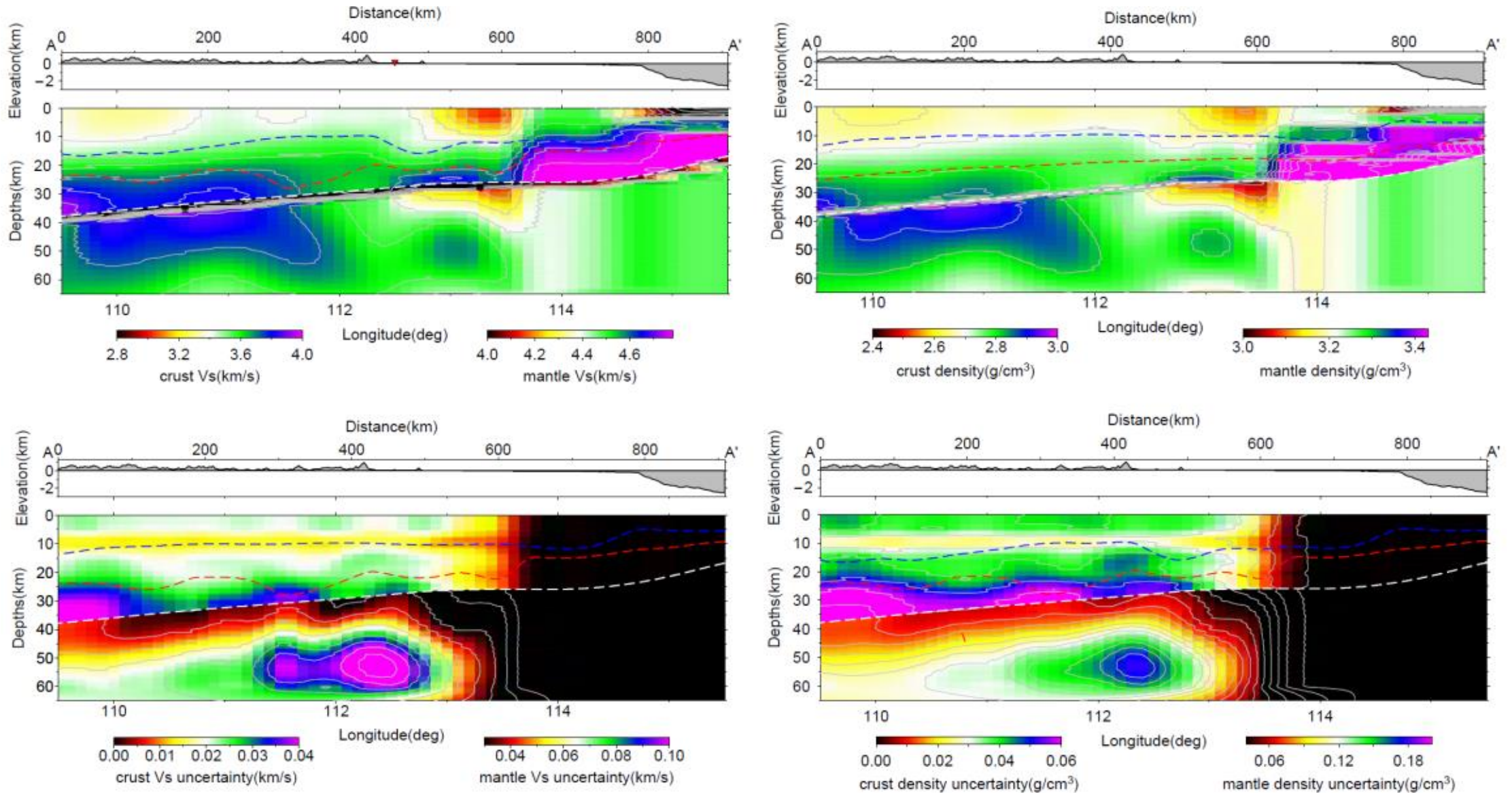

***Figure 3.*** *Cross-sections for shear-wave velocity (a) and density (b) structures in NW-SE directions through the JUNO site (Fig. 1). (c) and (d) show the estimated uncertainties for shear-wave velocity and density structures, respectively. The black areas with nearly zero uncertainties in (c) and (d) mean no uncertainty estimations for these areas because there are no resolution of the seismic ambient noise tomography. The dashed blue lines show the interface between the upper crust and mid-crust. The red dashed lines show the interface between the mid-crust and lower crust. The white dashed lines are the Moho discontinuities from joint constraints from gravity and deep seismic soundings (Hao et al., 2014).*

## 4. 3-D crustal model of U and Th abundances

Following the geophysical 3-D crustal model above, here we attempt to construct a 3-D model for the chemical (U and Th) abundances. Uranium and thorium abundance data of over 3000 individual surface rock samples in the research area, determined by Inductively Coupled Plasma-Mass Spectrometry or wet-chemistry method, are compiled. The geophysical 3-D model has divided the crust into upper, middle and lower crust. In the chemical abundance model, we further divide the upper crust of the South China continent and continental margin (referred to as continental in the following discussion) into two sub-layers — a top layer that is mainly composed of sedimentary rocks and granite intrusions, and a bottom layer that consists of Precambrian basement rocks. The boundary between these two layers is difficult to be distinguished by seismic tomography because of their similarity in seismic velocities. Here we arbitrarily assign the uppermost 5 km to be the top layer of upper crust (Zhang et al., 2013). The abundances of U and Th in each layer are then statistically evaluated based on the average composition of representative rock samples.

Establishing a refined model of the U and Th abundances of the uppermost layer of the continental crust is critical for the crustal geoneutrino flux prediction and error estimation. The uppermost layer on average has significantly higher U and Th abundances and lower density (Rudnick and Gao, 2003), and locates closer to the detector than the other layers. These features indicate that the uppermost layer makes greater contribution to the geoneutrino signal than any other layer (Huang et al., 2013). In addition, the chemical composition of the uppermost crust is more heterogeneous than the deeper layers (Rudnick and Gao, 2003). Therefore, more detailed geochemical information is needed for the top layer in order to precisely construct the abundance model.

When mapping the distribution of U and Th within the uppermost layer of the continental crust, we divide the Earth's surface into $0.5^{\circ} \times 0.5^{\circ}$ tiles that are projected vertically into discrete volume cells. Each cell is assigned with respective U and Th abundances and associated errors that are calculated from the proportions of different geologic units and the composition of each unit. The geologic units are divided based on lithology, and the relative proportions of different units are estimated based on published geological maps of the South China Block and South China Sea (Ma et al., 2002; Pubellier et al., 2017). The identified major geologic units include granites, intraplate basalts, mid-ocean ridge basalts (MORB), siliciclastic sedimentary rocks, and carbonates. The U and Th abundances and uncertainties of different geologic units are determined based on the statistics of literature data.

For each cell, if more than 5 samples are available for a certain geologic unit, its expected U and Th abundances and associated errors are calculated from the available data of this geologic unit in the cell. However, if less than 5 data points are available for a geologic unit within a cell, its expected U and Th abundances and errors are calculated from all data of this geologic unit in the entire research area. Where the data points are less than 10, the mean and the standard deviation are reported as the expected abundance and uncertainty, respectively. When more than 10 samples are available, the expected abundances and uncertainties are derived from a normal or lognormal distribution fit. The Kolmogorov-Smirnov (K-S) statistical test is applied to discriminate the normal and lognormal distributions.

Unlike the uppermost layer, the upper crust basement, middle crust, and lower crust of the continent generally show less variations in chemical compositions. Moreover, these deeper layers are less readily accessible for sampling. Here we use mid-grade metamorphic rocks and terranes, amphibolite facies xenoliths, and granulite facies xenoliths to represent the upper crust basement, middle crust and lower crust, respectively.

The results of our geochemical 3-D model are presented in Table 1 and shown in Fig. 4. Granites have higher average U ($5.4^{+5.6}_{-2.7}$ ppm) and Th ($23.7^{+16.2}_{-9.6}$ ppm) abundances than other lithologies (1σ). Intraplate basalts have similar U and Th abundances as oceanic basalts in South China Sea. Siliciclastic sedimentary rocks generally have higher U ($2.7^{+1.2}_{-0.8}$ ppm) and Th ($11.1^{+12.8}_{-5.9}$ ppm) abundances than carbonates (U = $0.9^{+0.5}_{-0.3}$ ppm, Th = $3.6^{+2.9}_{-1.6}$ ppm). Low-middle grade metamorphic rocks show relatively similar U and Th abundances, while their average U abundances decrease slightly with higher metamorphic grade. The average U abundances of slate, phyllite, schist and gneiss are 3.7±2.6 ppm, 3.5±1.5 ppm, $2.2^{+0.3}_{-0.8}$ ppm, and $2.5^{+2.3}_{-1.2}$ ppm, respectively.

The U abundances of the calculated surface cells vary from <1 to ~10 ppm, whereas the Th abundances of the surface cells vary from 4 to 20 ppm, with overall averages of U = 3.3 ppm and Th = 12 ppm. The average composition of low-middle grade metamorphic rocks is adopted as the model composition for the basement layer of continental crust, which has $3.0^{+2.1}_{-1.2}$ ppm U and $15.4^{+10.7}_{-6.3}$ ppm Th. Our results show that the U and Th abundances of the regional upper continental crust are higher than the averages of global continental crust (U = 2.7 ppm, Th = 10.5 ppm; Rudnick and Gao, 2003). This is consistent with the widely-distributed granite intrusions in this region, which are characterized by high HPE abundances.

The U and Th abundances adopted for the middle and lower continental crust are derived from the mean values of amphibolite facies xenoliths and granulite facies xenoliths in the research area. The U and Th abundances assigned for the middle continental crust are 0.6±0.4 ppm and 1.9±1.4 ppm, respectively, whereas those values of the lower continental crust are $0.04^{+0.08}_{-0.03}$ ppm and $0.1^{+0.3}_{-0.08}$ ppm, respectively. The U and Th abundances of these deeper layers of South China continental crust are lower than the global averages (Rudnick and Gao, 2003). Such distribution of U and Th abundances in the crust of South China Block is most likely due to extensive crustal reworking during its evolutionary history.

The oceanic crust is considered to be relatively homogeneous, and therefore we do not further divide the oceanic crust in South China Sea basin into layers. The composition of the oceanic crust can be represented and estimated by the composition of basalts sampled in the South China Sea basin. The method of statistical analyses is the same as the method described above for the surface continental crust layer. Our study suggests that the oceanic crust of South China Sea on average has U and Th abundances of $0.07^{+0.08}_{-0.04}$ ppm and $0.21^{+0.08}_{-0.03}$ ppm. This is consistent with the global average of oceanic crust (e.g., Sun and McDonough, 1989).

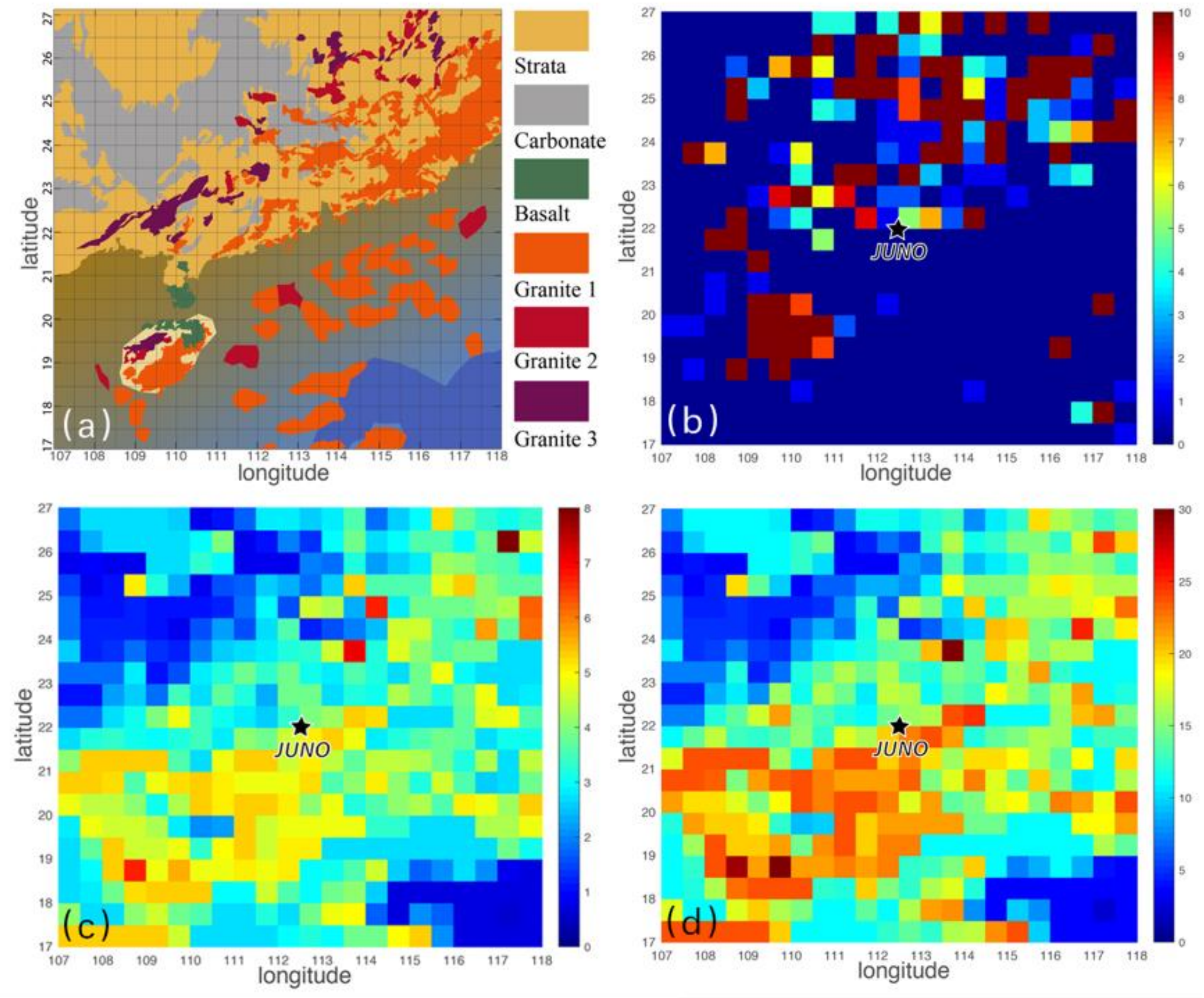

***Figure 4.*** *(a) geological map of the research area and a sketch of the 0.5 ° × 0.5 ° tiles in the surface layer of the geochemical 3-D model, where Strata = Sedimentary strata that is dominated by siliciclastic rocks, Granite 1 = Jurassic-Cretaceous granites, Granite 2 = Triassic granite, Granite 3 = Early Paleozoic granite. (b) available data points in each tile. (c) calculated U abundance of each tile. (d) calculated Th abundance of each tile.*

***Table 1.*** *U and Th abundances in geologic units in the regional crust in the 3-D model. N is the number of samples.*

| | | 1-sigma | | | | 1-sigma | | |
|---|---|---|---|---|---|---|---|---|
| **Geologic Unit** | **U mean** | **+** | **-** | **N** | **Th mean** | **+** | **-** | **N** |
| Granite | 5.4 | 5.6 | 2.7 | 1551 | 23.7 | 16.2 | 9.6 | 1551 |
| Intra-plate basalt | 0.7 | 0.7 | 0.5 | 255 | 3.7 | 3.2 | 1.8 | 255 |
| MORB | 0.7 | 0.8 | 0.4 | 332 | 2.6 | 1.8 | 1.2 | 332 |
| Siliciclastic rocks | 2.7 | 1.2 | 0.8 | 391 | 11.1 | 12.8 | 5.9 | 391 |
| Carbonats | 0.9 | 0.5 | 0.3 | 31 | 3.6 | 2.9 | 1.6 | 31 |
| Upper crust basement | 3.0 | 2.1 | 1.2 | 57 | 15.4 | 10.7 | 6.3 | 57 |
| Middle crust | 0.6 | 0.4 | 0.4 | 10 | 1.9 | 1.4 | 1.4 | 10 |
| Lower crust | 0.04 | 0.08 | 0.03 | 55 | 0.1 | 0.3 | 0.08 | 55 |
| Oceanic crust | 0.7 | 0.8 | 0.4 | 332 | 2.6 | 1.8 | 1.2 | 332 |

## 5. The Geoneutrino signal at JUNO

With the geophysical and geochemical 3-D crustal models conducted around the JUNO, the goal of this section is to predict the geoneutrino signal that would be possibly detected at the JUNO detector. When calculating the geoneutrino flux produced by the crust of the research area, we divide the regional crust into 4 layers, where each layer is further divided into 0.5 °× 0.5 °cells. The geophysical and geochemical features of each layer/cell are the same as discussed in previous sections. The total local crust contribution to the geoneutrino signal at JUNO can be obtained by taking the summation of the signal of each cell together layer by layer. The details of the mathematics of geoneutrino signal calculation is present in Appendix B.

Another important issue for the geoneutrino signal calculation is the evaluation of the associated uncertainty of the prediction. The Uncertainties of the estimated geoneutrino signal are mainly introduced by the uncertainties of the crustal model inputs (i.e., crust layer thickness, density and U-Th abundances), the cross section of the geoneutrino detection and the electron antineutrino survival probability. The latter two are expected to introduce uncertainties of about one percent to the geoneutrino signal (Han et al., 2019), which is not included in the uncertainty estimation in this study.

The uncertainty associated with the estimated crustal layer thickness is also minor. The thickness of crustal layers computed from seismic ambient noise tomography usually has less than 5% uncertainties (e.g., Christensen & Mooney, 1995). Our calculations show that if each crustal layer's thickness is under- or over-estimated by 5%, the total geoneutrino signal will correspondingly be under- or over-estimated by about 3%. The actual uncertainty propagated from the crustal thickness is expected to be even less significant (less than 1%), as the uncertainties from different units can be cancelled out during the calculation. This is true when the uncertainties of different units are considered to be uncorrelated, which is the case when the crustal layer thickness is obtained from the seismic ambient noise tomography method. Therefore, the contribution of uncertainties from the estimated crustal layer thickness is also negligible.

The method of Monte Carlo sampling is employed to construct the uncertainty of the geoneutrino signal associated with density and U-Th abundances. We conducted 1000 random sampling according to the statistical distribution of the input parameters, namely the density and U-Th abundances of each cell, and computed the corresponding geoneutrino signal 1000 times. The uncertainties associated with the input random numbers are propagated through the geoneutrino signal calculation to the distribution of the geoneutrino signal, which are characterized by the mean values and 1 $\sigma$ errors as reported in Table 2. Overall, the uncertainties of the top layer are significantly smaller than the other 3 layers. This is because of the statistically independent summation of the geoneutrino contributions from each cell of the top layer. In this case, one has the following two assumptions: Firstly, the rock samples have no correlation between different cells. Secondly, there are enough rock samples in each cell to describe the abundance distribution. The above assumptions are not always true in our model, since some cells do not have enough samples so that the composition of those cells can be independently constrained. Therefore, the error of the estimated geoneutrino flux presented here is the possible minimal error. The total prediction from the refined Local Crust is $S_{LOC}$=28.5 ± 4.5 TNU (a Terrestrial Neutrino Unit (TNU) is one geoneutrino event per $10^{32}$ target protons per year).

***Table2** the predicted geoneutrino signal(S) from local crust in the 10°x10° area around JUNO and uncertainty(σ) in TNU, the results include the contribution from Continental Crust and Oceanic Curst, the geoneutrino signal spectrum are in Fig. B1 in Appendix B.*

| | | $S_U \pm \sigma$ | $S_{Th} \pm \sigma$ | $S_{U+Th} \pm \sigma$ |
|---|---|---|---|---|
| Upper Crust | Top Layer | $10.5^{+0.7}_{-0.7}$ | $3.2^{+0.3}_{-0.3}$ | $13.8^{+0.8}_{-0.7}$ |
| | Basement | $8.1^{+7.0}_{-3.7}$ | $2.6^{+1.8}_{-1.1}$ | $11.0^{+5.9}_{-3.9}$ |
| Middle Crust | | $1.7 \pm 1.0$ | $0.4 \pm 0.3$ | $2.1 \pm 1.1$ |
| Lower Crust | | $1.9^{+3.8}_{-1.3}$ | $0.8^{+5.7}_{-0.7}$ | $1.7^{+4.0}_{-1.2}$ |
| Oceanic Crust | | $0.2 \pm 0.05$ | $0.1 \pm 0.01$ | $0.3 \pm 0.05$ |
| Total | | $21.3 \pm 4.0$ | $6.6 \pm 1.3$ | $28.5 \pm 4.5$ |

To constrain the geoneutrino contribution of the Far Field Crust (FFC) outside the selected 10°x10° area around the JUNO detector, we use the density and thickness of the global CRUST1.0 model (Laske et. al., 2013) as the input. The abundances of U and Th for the Upper and Middle Crust are adopted from Rudnick and Gao (2003). For the Lower Crust, values in the literatures span a rather large range, corresponding to different assumptions about the relative proportion of mafic and felsic rocks. Here we employ a mean value together with an uncertainty to account for the spread of published values. For sediments, we follow the work of Plank et al. (1998). As a result, calculation shows that the Far Field contribution of geoneutrino is estimated to be $S_{FFC} = 9.8 \pm 1.7$ TNU.

The bulk crust geoneutrino signal expected at JUNO, corresponding to 38.3±4.8 TNU, can be expressed as the sum of two independent contributions, the signal from the local 10°x 10° crust surrounding JUNO (LOC) and the signal from the rest of the global crust (FFC). The local crust contributes around 74% of the total crust geoneutrino signal, among which 66% is from the upper crust. The geoneutrino signal from the CLM (Continental Lithospheric Mantle), DM (Depleted Mantle) and EM (Enriched Mantle) are inherited from Strati et al. (2015). Consequently, we can accomplish the total geoneutrino signal prediction at JUNO (Table 3). In Fig. 5, the cumulative crust geoneutrino signal and its percentage in total geoneutrino signal are plotted versus the distance to the JUNO site, from which one can see that the local 10 ° × 10 °area contributes to more than 55% of the total signal.

***Table 3.*** *Summary of geoneutrino signals and their uncertainties (1 standard deviation) in TNU expected for JUNO, from Uranium ($S_U$), Thorium ($S_{Th}$), and their sum ($S_{TOT}$). The Bulk Crust contribution includes that of Local Crust (LOC) and Far Field Crust (FFC). Lithosphere contribution is the sum of the Bulk crust and Continental Lithospheric Mantle (CLM) contributions. The total signal is the sum of that of Lithosphere, Depleted Mantle (DM), and Enriched Mantle (EM).*

| | $S_U \pm \sigma$ | $S_{Th} \pm \sigma$ | $S_{TOT} \pm \sigma$ |
|---|---|---|---|
| **LOC** | 21.4±4.0 | 6.5±1.3 | 28.5±4.5 |
| **FFC** | 7.6±1.6 | 2.3±0.2 | 9.8±1.7 |
| **Bulk Crust** | 29.0±4.3 | 8.8±1.3 | 38.3±4.8 |
| **CLM** | $1.3^{+2.4}_{-0.9}$ | $0.4^{+1.0}_{-0.3}$ | $2.1^{+3.0}_{-1.3}$ |
| **Lithosphere** | $30.3^{+4.9}_{-4.4}$ | $9.2^{+1.7}_{-1.4}$ | $40.4^{+5.6}_{-5.0}$ |
| **DM** | 4.2 | 0.8 | 4.9 |
| **EM** | 2.9 | 0.9 | 3.8 |
| **Total** | $37.4^{+4.9}_{-4.4}$ | $10.9^{+1.7}_{-1.4}$ | $49.1^{+5.6}_{-5.0}$ |

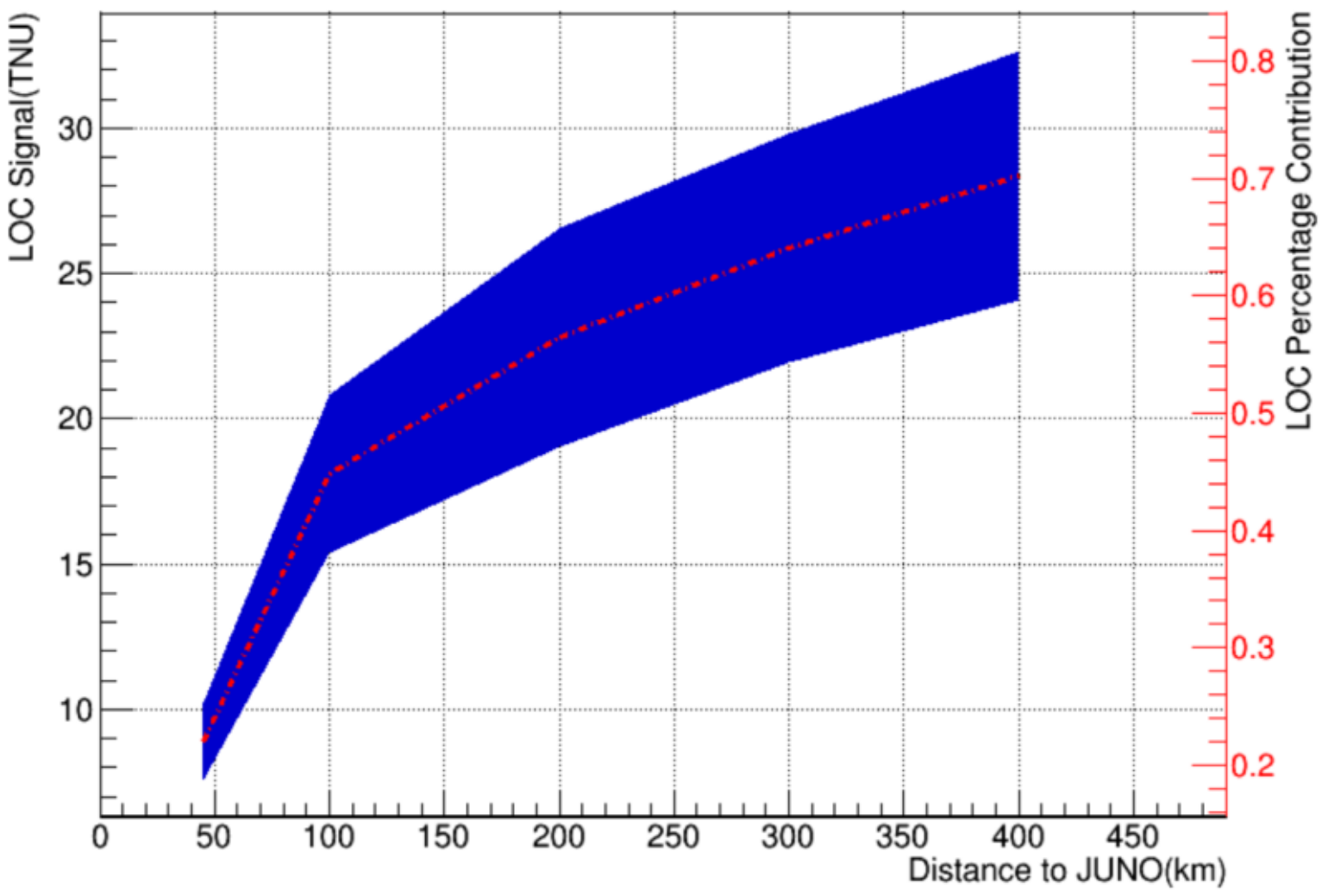

***Figure 5.*** *The distribution of local crust geoneutrino signal and its contribution to the total lithosphere geoneutrino signal at different straight-line distance to JUNO detector. The Blue range are the local geoneutrino signal and expected error bar, the red dotted line is the local percentage contribution to the total lithosphere geoneutrino signal.*

The geoneutrino signal depends on both the total mass of HPEs in the Earth's interior and the geochemical and geophysical properties of the region around the detector. The combination of this information with the regional study allows us to build up the connection between the geoneutrino signal and the masses of HPEs (or radiogenic heat) inside the Earth (see the detailed discussion in Appendix C). As a result, the relation of expected geoneutrino signal from U at JUNO with the U mass (m(U)) radiogenic heat is illustrated in the left panel of Fig. 6, where the signal uncertainty $\delta(U)$ is defined by the range between the high and low signals marked as $S_{high}$ and $S_{low}$, respectively. In general, the calculation can be developed individually for U and Th, with the connections between the geo-neutrino signals S(U) and S(Th) and their respective masses m(U) and m(Th) inside the Earth respectively. However, it is rather difficult to separate the geo-neutrino signals of U and Th from the current experimental measurements. We expect that the next-generation geoneutrino detectors like JUNO with high energy resolution would be able to collect sufficiently enough events in order to distinguish the U and Th signals from each other (Han et al., 2016). Currently, we rely on the chondritic estimate of the Th/U ratio in order to derive global abundances of these two elements. The predicted total geoneutrino signal at JUNO S(U+Th) as a function of the radiogenic heat $H_R$(U+Th) inside the Earth is plotted in the right panel of Fig. 6.

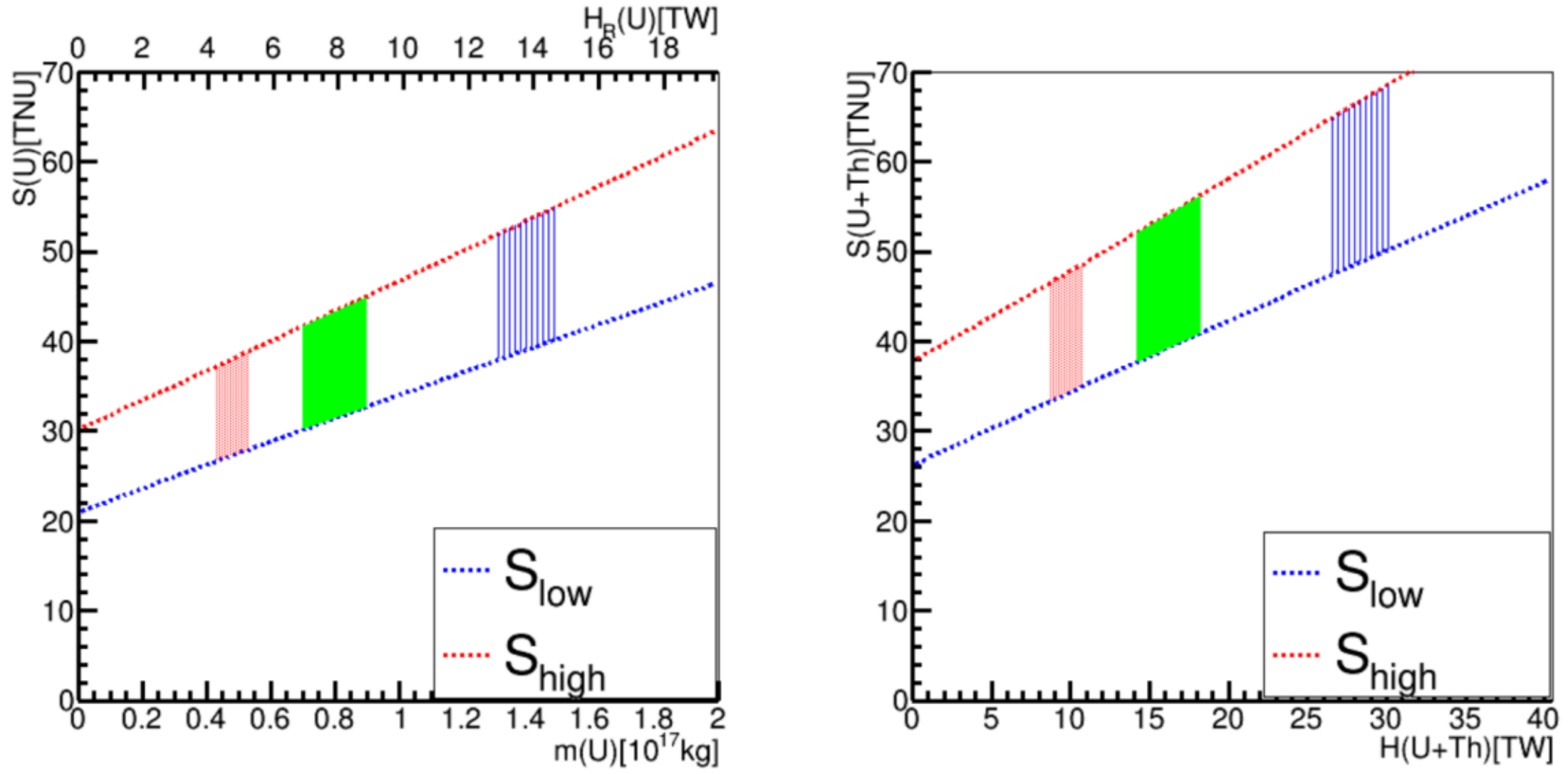

***Figure 6.*** *The connection between the geoneutrino signal and the masses of HPEs (or radiogenic heat) inside the Earth, derivative from the combination information with the regional study and different mantle models. Left plot: the predicted geoneutrino signal from U with the mass of U and the radiogenic heat production. Right plot: predictions on the combined signal S(U+Th) at JUNO as a function of the radiogenic heat production rate H(U+Th). The range between $S_{high}$ and $S_{low}$ illustrate the geoneutrino signal uncertainty $\delta(U)$. From left to right, the quadrangle regions between the red and blue lines denote the regions allowed by cosmochemical model (expected m(U) at the range of 0.7~0.9 ×$10^{17}$kg), geochemical model(expected m(U) at the range of 0.43~0.53 ×$10^{17}$kg) and geodynamical model(expected m(U) at the range of 1.3~1.5 ×$10^{17}$kg), respectively.*

## 6. Discussion

### *6.1 The Local 3-D Geophysical Model*

Compared to the earthquake-based travel time or surface wave tomography, whose resolutions are largely limited by the distribution of earthquakes, the seismic ambient noise tomography method is not limited by the distribution of earthquakes and has higher and relatively uniform resolution on the crust beneath the seismic array (Shapiro et al., 2005; Yang et al., 2011; Shen et al., 2016; Li et al, 2016). Although the deep seismic soundings with active sources have higher resolution of the crust, the spatial coverage is strongly limited by the high-cost. We take the results of deep seismic soundings as the prior constraints on the Moho depth (Hao et al., 2014), and constructed the local 3-D geophysical model by the seismic ambient noise tomography method, which generates a higher and more uniform spatial resolution for the whole terrane of the South China Block. The results in this study are largely consistent with the previous model with ambient noise tomography in China (Shen et al., 2016), but our results focus in the South China Block under more constraints of Moho depth with deep seismic soundings, which can ensure a more reliable model.

The depths of the boundaries between the upper-mid crust and mid-lower crust are largely comparable between the models from this study and those of CRUST1.0, especially in the South China Block (Fig. A6 in Appendix A). The results from this study show more detailed tectonic characteristics in the crust, demonstrating higher resolution than the CRUST1.0 model. For most areas of the South China Block, the boundary depths are about 10~20 km for the upper-mid crust, and about 20~25 km for the mid-lower crust. Howbeit in the eastern margin of Tibet Plateau, the depths of the upper-mid crust boundary can be up to 40 km, which can be attributed to the improper isosurfaces of P-wave velocities of 6.0 km/s for this area in contrast to the relatively low velocity crust in the eastern margin of Tibet Plateau. Since the density model is derived from the 3-D velocity model based on empirical relations between the density and velocity, the 3-D density model exhibits characteristics similar to the velocity model. The South China Block shows high-velocity and high-density anomalies in the upper and middle crust at 5~25 km. At lower crust depth (~30 km), relatively low-velocity and low-density anomalies appear when compared to the South China Sea. To the uppermost mantle at 35~45 km depth, significant high-velocity and high-density anomalies are observed beneath the South China Block. In contrast, the sedimentary basins (e.g., Sichuan Basin, Jianghan Basin and North China Basin) show low-velocity and low-density anomalies at the depths shallower than ~ 10 km, whereas the eastern margin of Tibet Plateau has low-velocity and low-density anomalies in the mid-lower crust at depths of 20~45 km.

*6.2 The Geology and Geochemical 3-D Model*

The predicted bulk crustal geoneutrino signal at JUNO (38.3±4.8 TNU) is somewhat higher (~23%) than those predicted for other observation stations (e.g., $31.1^{+8.0}_{-4.5}$ TNU for SNO+, Strati et al., 2017). The estimated geoneutrino flux at JUNO in this study is also significantly higher (~36%) than previous studies (e.g., $28.2^{+5.2}_{-4.5}$ TNU by Strati et al., 2015) using the global CRUST1.0 model. The high geoneutrino flux is the result of the relatively high average U and Th abundances of the upper crust in our geochemical model. This is consistent with the widespread distribution of high U-Th granite intrusions in the area around JUNO, which is also reflected by relatively high heat flow in this region (Wang, 2001).

There have been contrasting models for the evolution of the South China crust. Some researchers suggest that the massive granite intrusions are primarily the result of reworking and partial melting of the lower crust (e.g., Wang et al., 2006; Huang et al., 2011). Others proposed that mantle-derived melts have induced partial melting of the lower crust and thus generated the granite rocks (e.g., Li et al., 2005; Yu et al., 2016). In the first case, melt extraction from the lower crust would take away incompatible elements (such as U and Th) and transport them to the upper crust, leaving lower crust residues with low U and Th abundances. In contrast, for the second case, infiltration of mantle-derived melts to the lower crust would add significant amounts of incompatible elements to the lower crust, and thus increase the average U and Th abundances of the lower crust.

Our geochemical model suggests significantly lower U and Th abundances for the lower crust of the South China than global average, in which the constraints mostly come from measured geochemical compositions of local granulite xenoliths. If these granulite samples are representative for the local lower crust, the low U and Th abundances of the lower crust would support partial melting of lower crust without significant material inputs from the mantle. Therefore, our model supports recent reworking of the South China crust with limited mantle-derived melts to account for the massive granite formation in this region.

*6.3 The Future Precision for the JUNO measurement*

To discriminate different models of mantle composition with the JUNO geoneutrino measurement, the uncertainty of the estimated local crust signal needs to be reduced to less than 8% in order to reach a significance of better than 3σ for the mantle signal estimate (Han et al., 2016). In this work, the local crust contribution of the geoneutrino signal is 28.5±4.5 TNU, corresponding to an uncertainty of 15.7%, which is better than the global model of 17% but still needs further improvement. Our calculation suggests that both the basement layer and top layer of the local crust are the main sources of geoneutrinos. However, the uncertainty from the top layer is significantly lower than that from the basement layer (Table 2). One important reason is the tremendous difference in the sample quantities of the two layers. For the top layer, we assume that the composition of each cell is independently constrained, and thus there is no correlation between the errors of each cell. Consequently, the uncertainty of the top layer is reduced to be around 5%. In contrast, the U and Th abundances of the whole basement layer are estimated based on the composition of low-grade metamorphic rocks, and considered to be uniform. Thus, the errors of different cells are correlated, resulting in an uncertainty as large as 45%. Therefore, the apparent solution to reduce the uncertainty is to obtain additional samples of the basement layer to reduce the uncertainty correlation between different cells. Moreover, the uncertainty also depends on the size of each cell. The smaller the cell size is, which means there are more independent cell units, the smaller the uncertainty will be, as shown Fig. B2 in Appendix B.

In the near future, we plan to update the local crustal model by collecting more geophysical and geochemistry data, especially for the basement layer. The boundary between the top layer and the basement layer is arbitrarily divided roughly at 5 km for the whole range in the current model. In future model, surface heat flow data can be used to constrain the actual boundary depths. With improved uncertainty for the estimation of local crust geoneutrino signal, JUNO is expected to be able to distinguish different mantle models. Furthermore, if the mantle contribution is similar at different locations, the combined observations from JUNO and other observatories (such as KamLAND, Borexino, SNO+) will help test different crustal models (Fiorentini et al., 2012).

## 7. Conclusions

In this paper, we have constructed a refined 3-D model of the local crust (10 ° × 10 °) surrounding JUNO, which will have the largest geoneutrino detector on Earth and is scheduled to be online in 2021. Geophysical crustal model via seismic ambient noise tomography indicates the thickness of the regional crust varies from ~40 km in the South China Block to <10 km in the South China Sea basin. For most areas of the South China Block, the boundary depths are about 10~20 km for the upper-mid crust, and about 20-25 km for the mid-lower crust. Our geochemical crustal model suggests the U and Th abundances of the upper crust of South China Block are higher than global average, with the topmost layer on average having U of ~3.3 ppm and Th of ~12 ppm and the basement layer on average having U = $3.0^{+2.1}_{-1.2}$ ppm and Th = $15.4^{+10.7}_{-6.3}$ ppm. In contrast, the middle and lower crust of the South China Block are predicted to have lower U and Th abundances than global average, with U = 0.6±0.4 ppm, Th = 1.9±1.4 ppm and U = $0.04^{+0.08}_{-0.03}$ ppm and Th = $0.1^{+0.3}_{-0.08}$ ppm, respectively. These features indicate that the recent massive granitic magmatism is most likely the result of extensive reworking and melting of the lower crust without significant inputs of mantle-derived melts.

The predicted geoneutrino signal at JUNO is 38.3±4.5 TNU on the basis of our refined 3-D crustal model, which is significantly higher than previous predictions. This is consistent with the widespread distribution of high U-Th granite intrusions in the area. In this work, the uncertainty of the predicted geoneutrino signal is ~15.7%, which is slightly better than that of the global model (~17%). Most of the uncertainty is inherited from the uncertainty of the geochemical model, and the contribution to the uncertainty from the top layer of upper crust is reduced from ~35% to ~6% by constructing the geochemistry properties cell by cell. Future work will continue to focus on improving the precision of experimental measurements and building more reliable crustal models, so that the different mantle models may be distinguished base on the JUNO observation.

**Acknowledgements:**

This work is supported by the National Key R&D Program of China (2018YFA0404100) and the Ministry of Education of the PRC funding (2-9-2017-214). Waveform data in South China for this study are provided by Data Management Centre of China National Seismic Network at Institute of Geophysics, China Earthquake Administration (doi:10.11998/SeisDmc/SN, http://www.seisdmc.ac.cn). Constructive reviews from Mr. Ivan Callegari and an anonymous reviewer are appreciated.

## Abbreviation

**HPEs**： Heat Producing Elements

**BSE**： Bulk Silicate Earth

**JUNO**: Jiangmen Underground Neutrino Observatory

**SNRs**: Signal-to-Noise Ratios

**NCFs**: Noise Cross-correlation Functions

**MORB**: Mid-Ocean Ridge Basalts

**K-S:** Kolmogorov-Smirnov

**UC_LOC**: LOCal Upper Crust

**MC_LOC**: LOCal Middle Crust

**LC_LOC**: LOCal Lower Crust

**UUC_LOC**: LOCal Upper Upper Crust

**LUC_LOC**: LOCal Lower Upper Crust

**FFC**: Far Field Crust

**CLM**: Continental Lithospheric Mantle

**DM**: Depleted Mantle

**EM**: Enriched Mantle

**IBD**: Inverse Beta Decay

## Appendix A: Extended Graphs

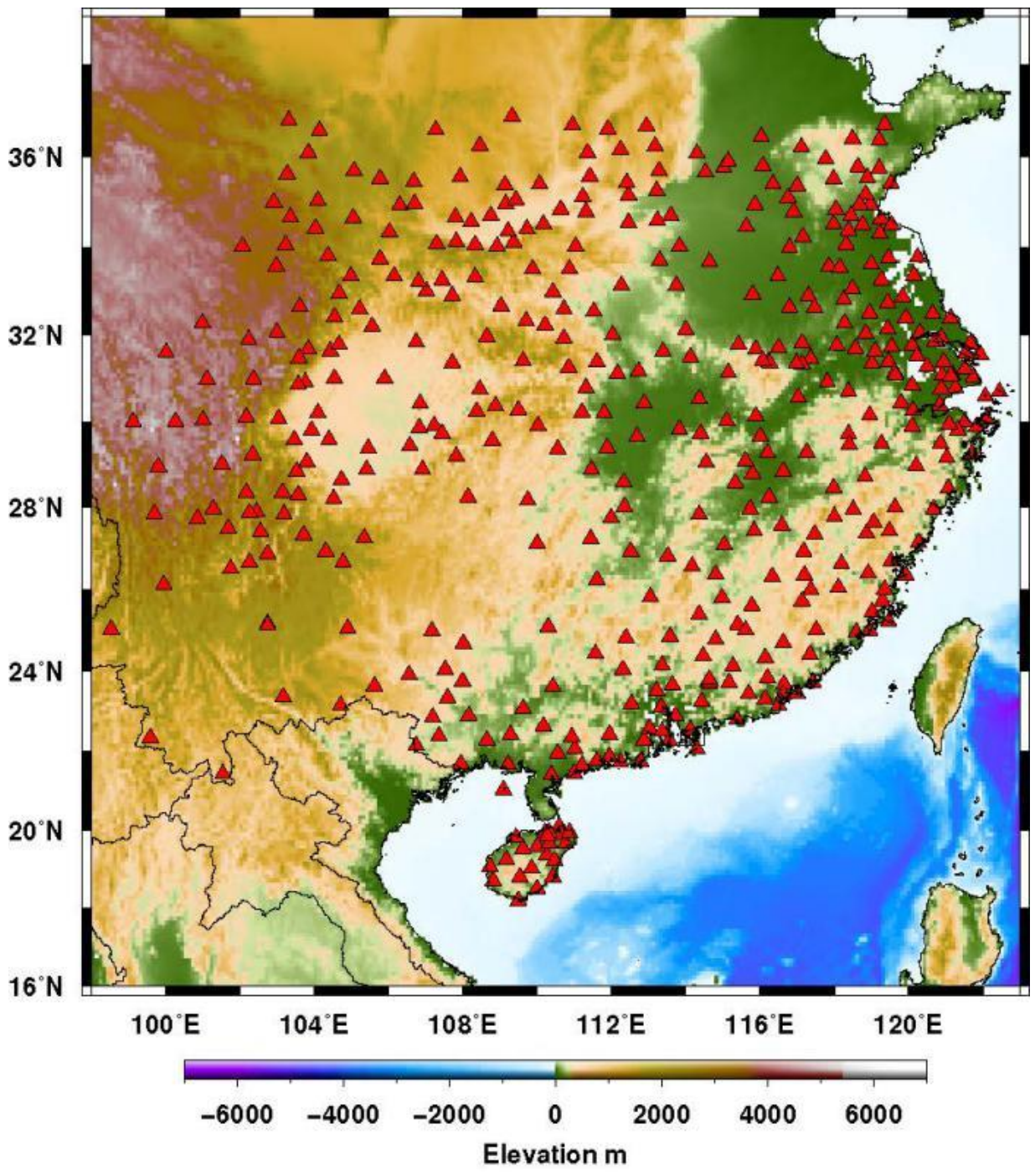

***Figure A1***. *Distributions of permanent seismic stations (red triangles) used in the seismic ambient noise tomography in the South China Block.*

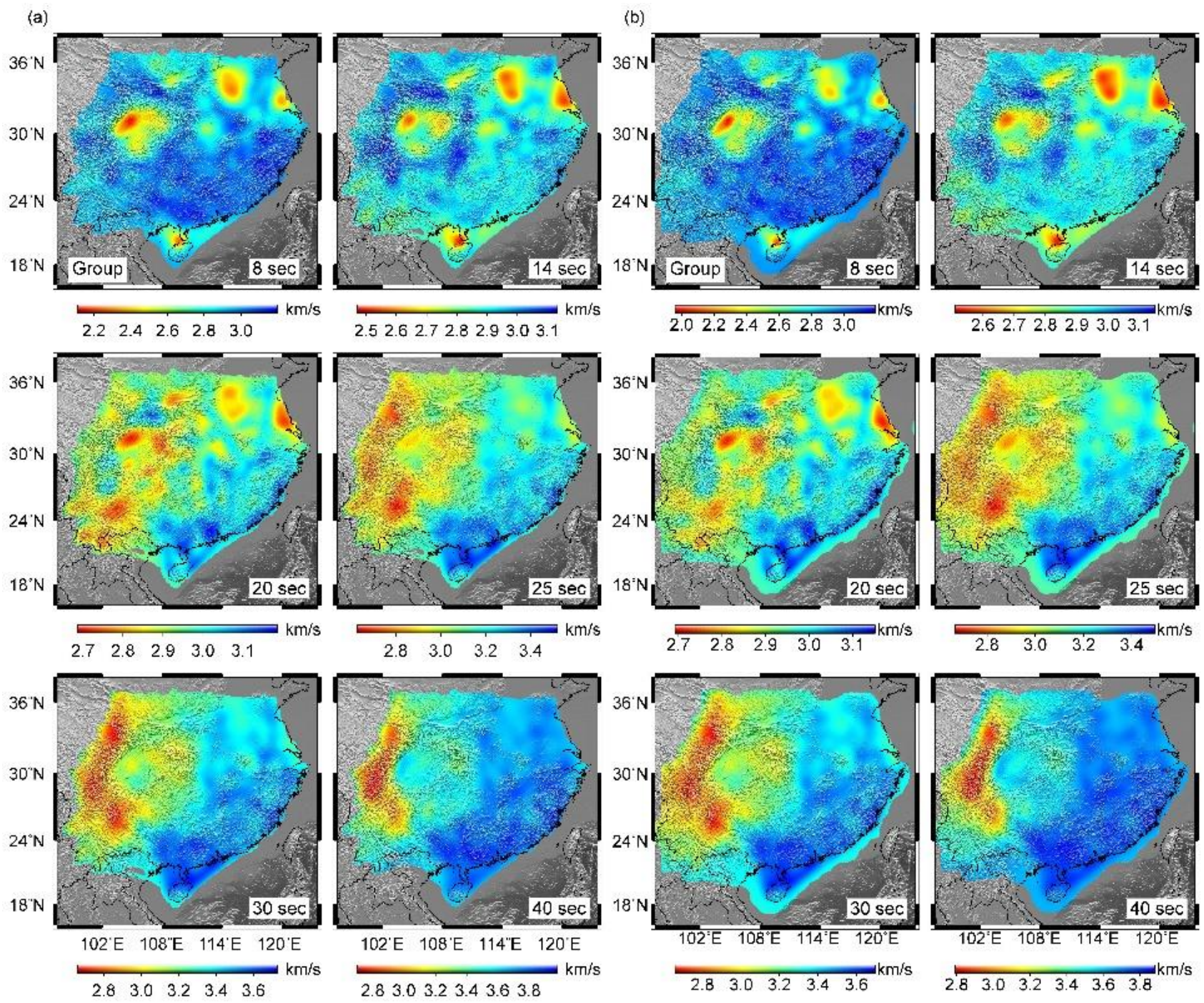

***Figure A2.*** *(a) Inverted group velocity distributions at 8~40 s periods from surface wave tomography in the South China Block. (b) Predict group velocity distributions at 8~40 s periods based on the inverted 3-D shear-wave velocity structures from seismic ambient noise tomography (Fig. A4 in Appendix A). The consistency between the inverted and predict group velocity maps at each corresponding period suggests the 3-D velocity model obtained in this*

*study can well fit the seismic observations.*

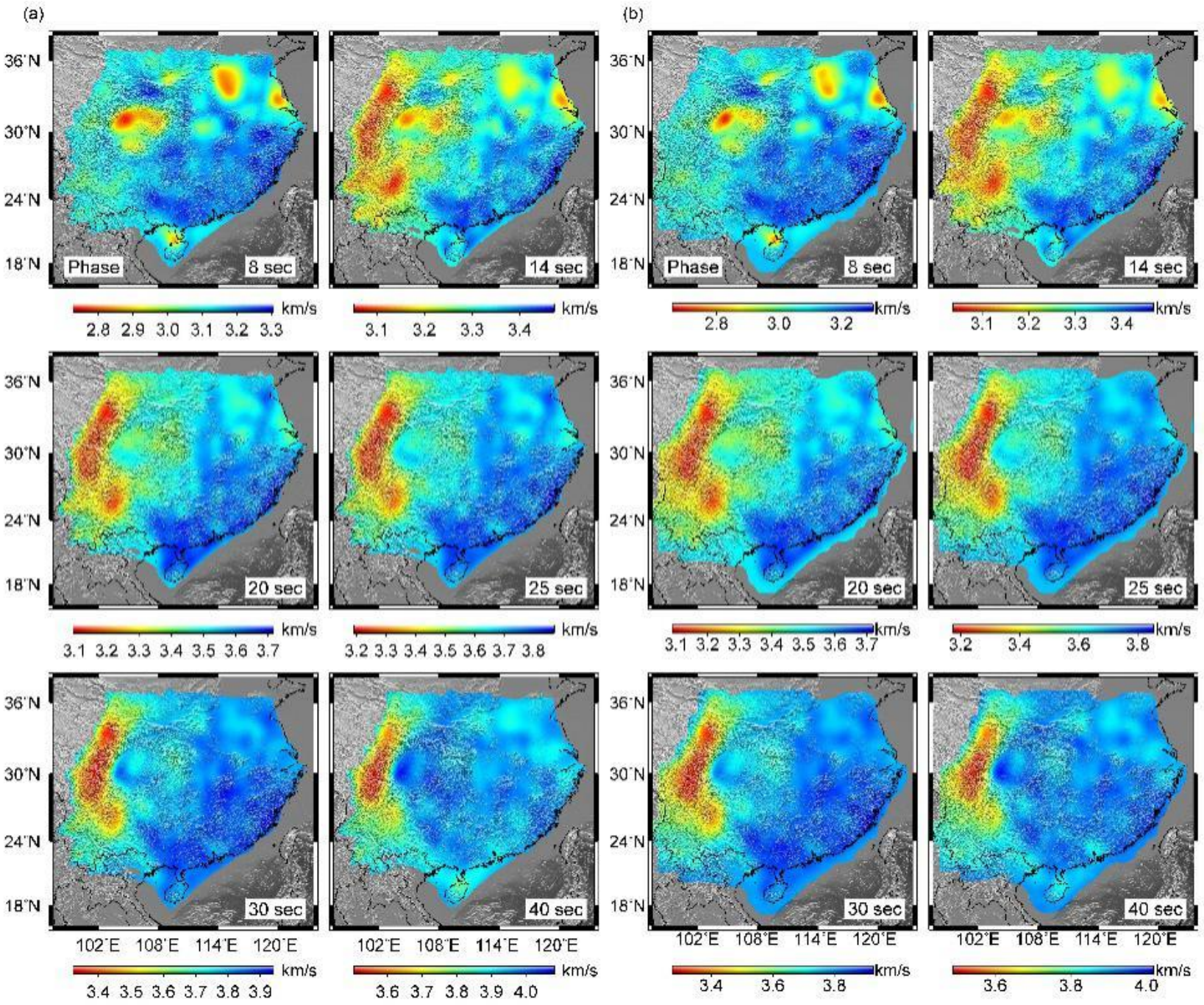

***Figure A3.*** *(a) Inverted phase velocity distributions at 8~40 s periods from surface wave tomography in the South China Block. (b) Predict phase velocity distributions at 8~40 s periods based on the inverted 3-D shear-wave velocity structures from seismic ambient noise tomography (Fig. A4 in Appendix A). The consistency between the inverted and predict group velocity maps at each corresponding period suggests the 3-D velocity model obtained in this study can well fit the seismic observations.*

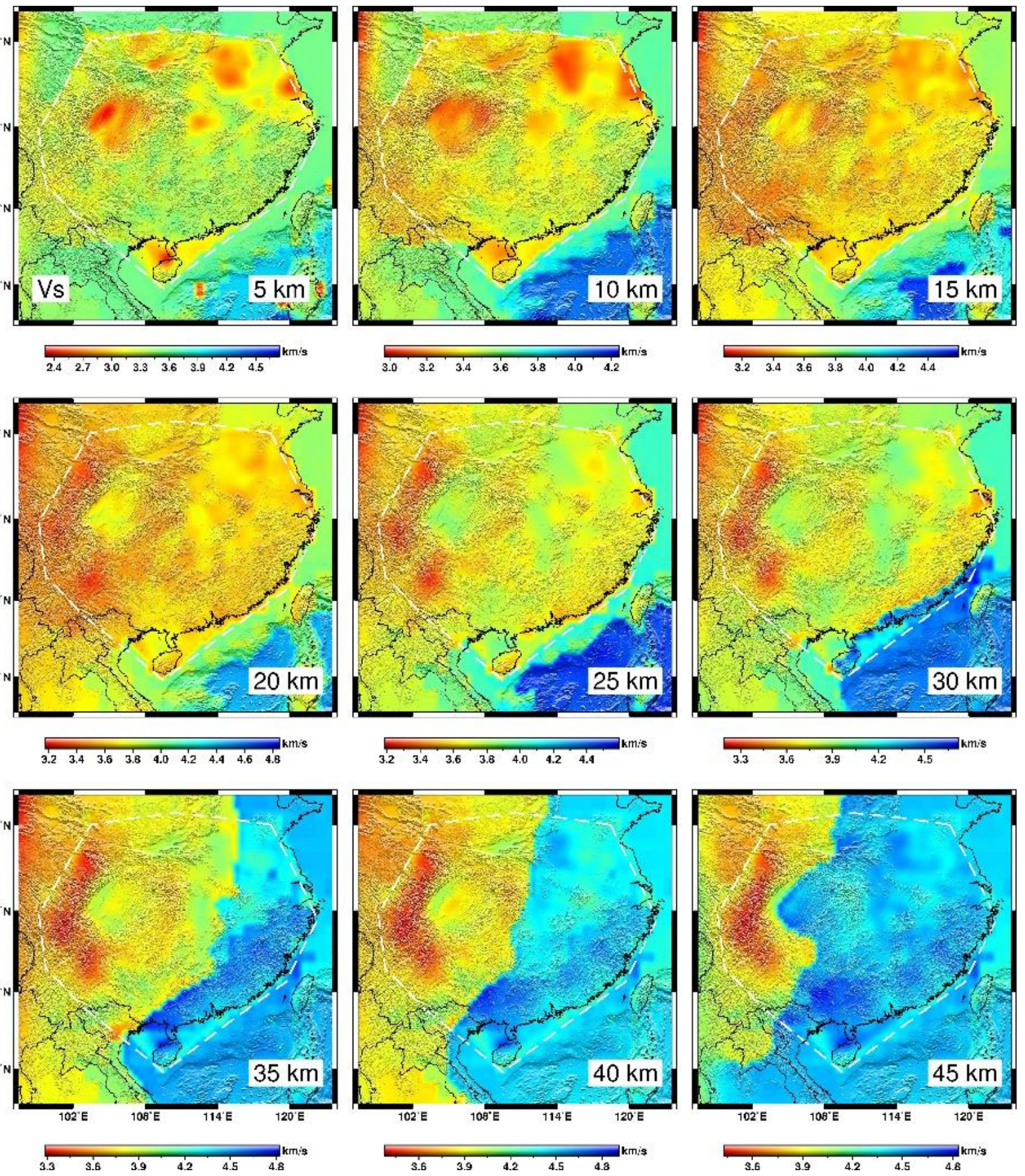

***Figure A4.*** *3-D shear-wave velocity distributions at 5~45 km depths from the seismic ambient surface wave tomography in the South China Block. The white dashed line shows the regions with high-resolution results derived from seismic ambient noise tomography. As for the outside regions, the structures are derived from the global CRUST1.0 model.*

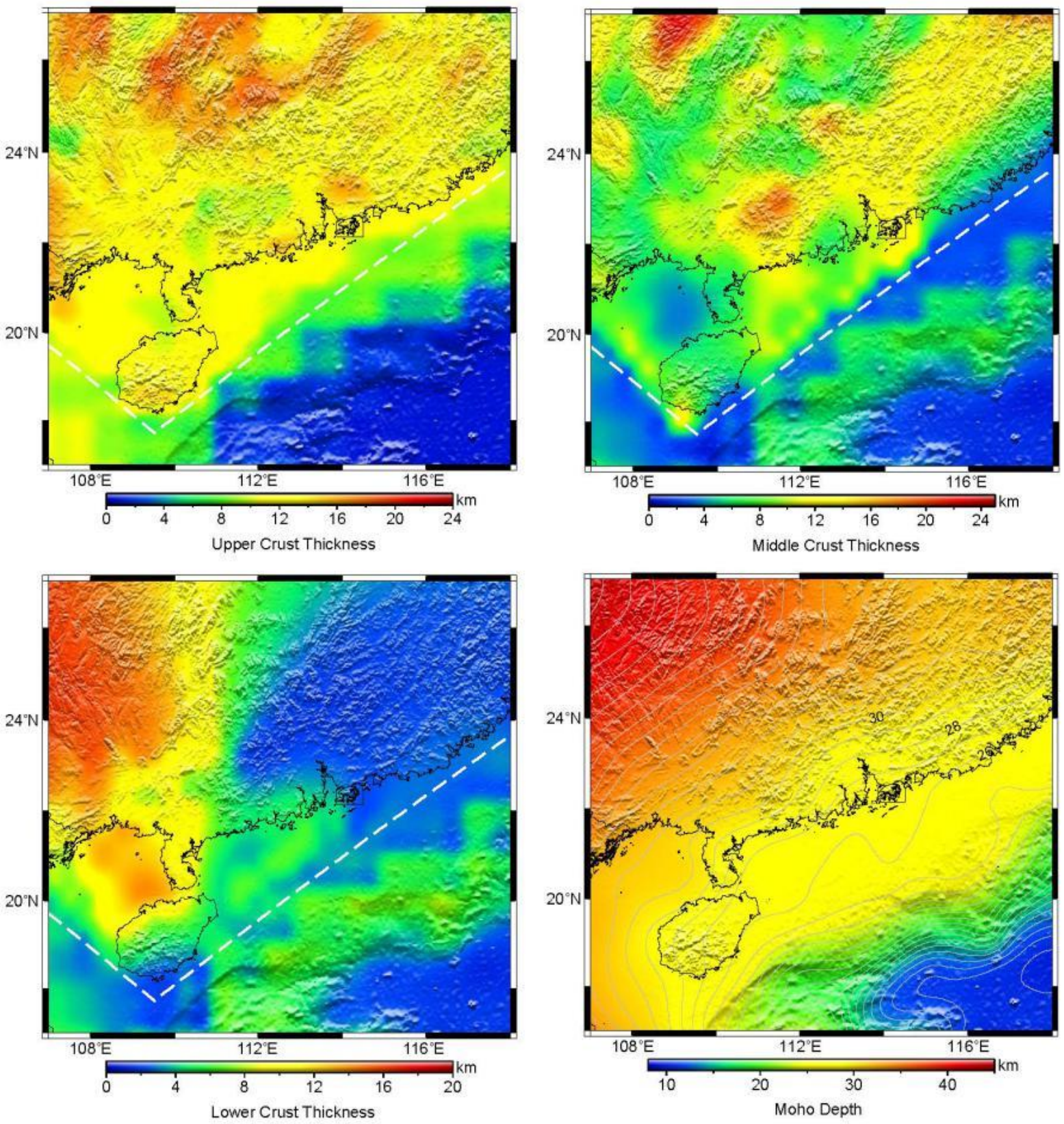

***Figure A5***. *(a-c) The thickness of the upper crust, middle crust and lower crust in the area surrounding the JUNO site. (d) The Moho depth distributions in the area surrounding the JUNO site (Hao et al., 2014). The white dashed line shows the regions with high-resolution results derived from seismic ambient noise tomography. As for the outside regions, the structures are derived from the global CRUST1.0 model.*

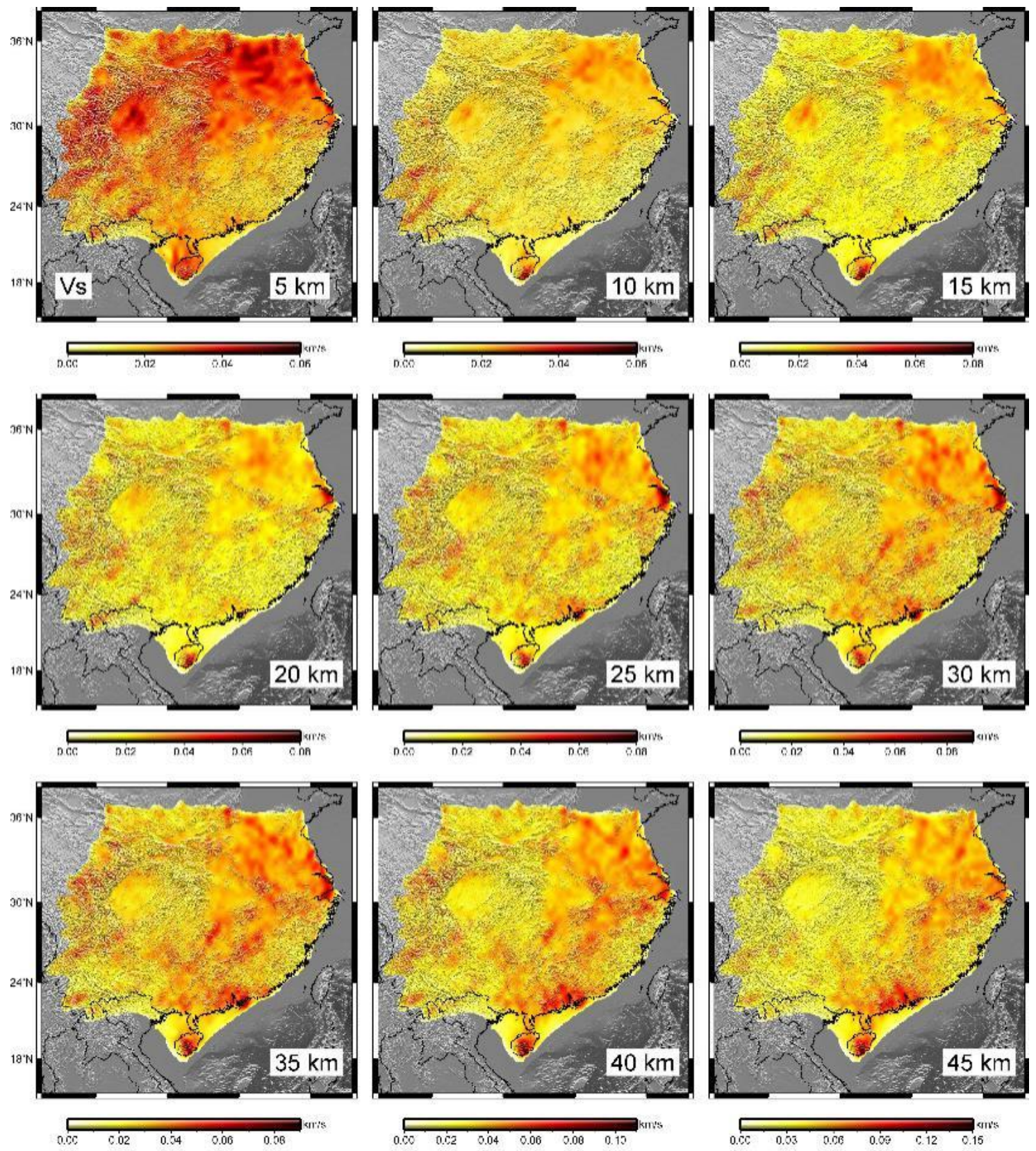

***Figure A6.*** *The uncertainties of the 3-D shear-wave velocity distributions at 5~45 km depths for the seismic ambient surface wave tomography estimated by the bootstrap method. The white dashed line shows the regions with high-resolution results derived from seismic ambient noise tomography. As for the outside regions, the structures are derived from the global CRUST1.0 model.*

## Appendix B: Expected Spectrum at JUNO and Uncertainty Study with Different Cell Size

The fully oscillated geoneutrino signal from each radionuclide at observation location is calculated from

$$N_i = N_p t \int dE \epsilon(E) \sigma(E) \frac{d\Phi}{dE}$$

$N_i(x)$ is the geoneutrino signal in TNU (Terrestrial Neutrino Unit) from isotope $i$, where TNU is the unit representing the number of geoneutrino events detected with one-year data taking and $10^{32}$ free protons, which is approximately similar to the event number of a 1 kton liquid scintillator detector. The parameter $\epsilon(E)$ is the detector efficiency but is assumed to be one in this calculation. The geoneutrino signal is detected via the Inverse Beta Decay (IBD) reaction, and the cross section $\sigma(E)$ is taken from Strumia and Vissani (2003). $\Phi$ is the geoneutrino flux from the interested crustal region, the relation between the flux and the geologic information is given as follows:

$$\frac{d\Phi(i)}{dE} = \int dV \frac{\rho(\vec{r})}{4\pi L^2} \frac{A(i,\vec{r})}{\tau_i m_i} P(E,L)$$

where $i$ is the considered radioactive isotope, and can be either U or Th, $\tau_i$ is the isotope life time, $m_i$ is the atomic mass. The spatial distribution of the rock density $\rho(\vec{r})$ and the abundance $A(i,\vec{r})$ in each cell are provided with the local refined reference model. $L$ is the distance from the reference cell to the detector and implicitly encodes the crustal structure of the refined geo-physical model. As for the survival probability, we use the baseline-averaged probability for simplicity, which approximates as $P_{ee} = 0.55$ (Capozzi et al, 2014). Finally, $f_i(E)$ is the geoneutrino energy spectrum from the considered isotope and is taken from Ref. (Araki et al. 2005). The expected geoneutrino spectrum at JUNO with 3%/√E energy resolution is shown in

Fig. B1 in Appendix B.

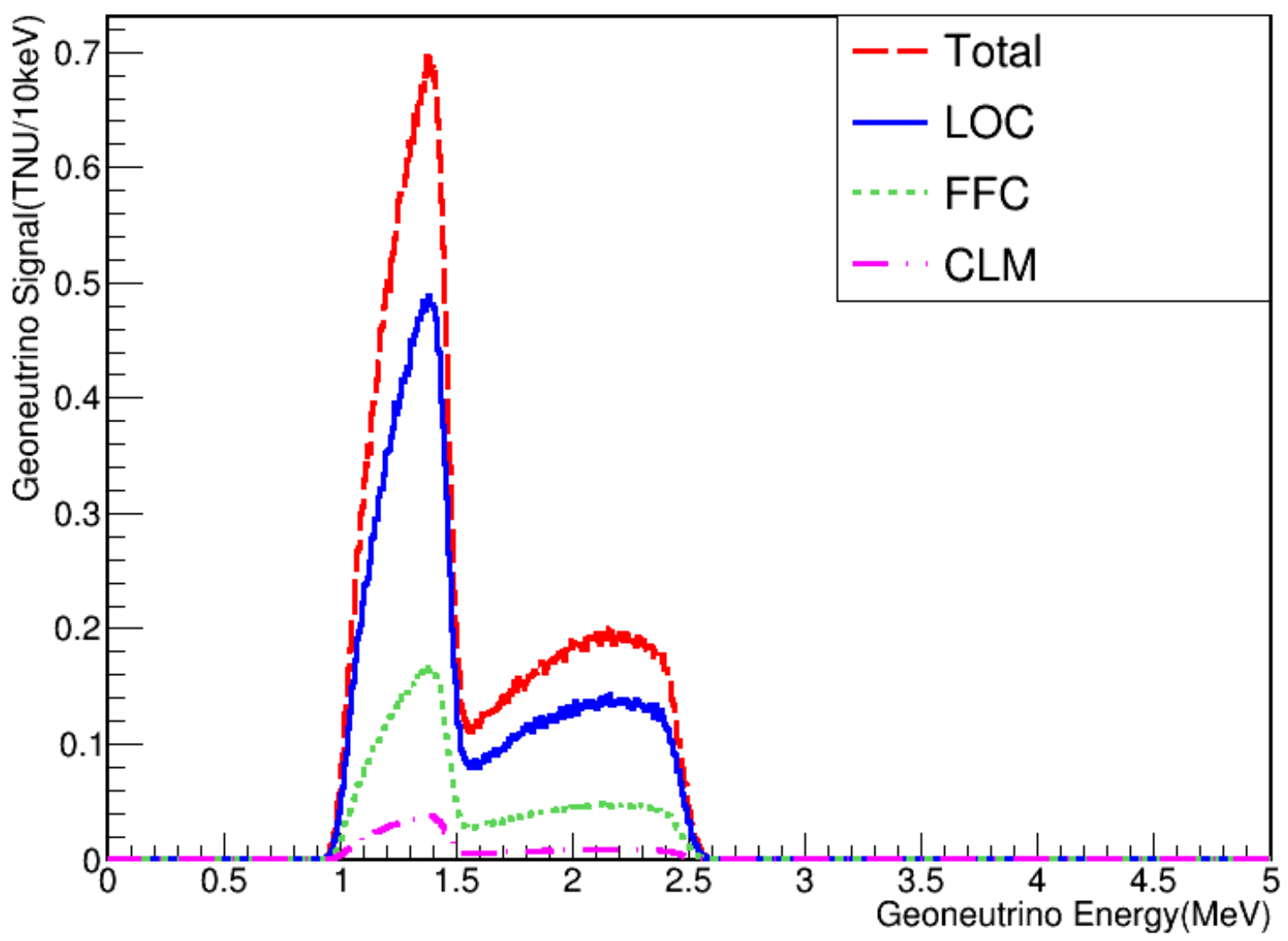

***Figure B1.*** *The predicted geoneutrino spectrum from different ranges at JUNO detector with 3%/√E resolution.*

We use the method described above to simulated one thousand possible JUNO crustal geoneutrino measurements using a Monte Carlo approach. In each simulation, we have attributed to each spectral component a rate randomly extracted from the Gaussian distributions according to CRUST1.0 global structure model and global crustal composition model (Rudnick and Gao, 2003). In the simulations we have included 5% thickness and density uncertainties each, and the uncertainties from the U/Th composition inherited from (Rudnick and Gao, 2003). The possible precision of crustal geoneutrino signal prediction under these conditions was evaluated for 1°×1° and 0.5°×0.5° cell size, we assumed the uncertainty has no correlation between different cells. Fig.B2 shows the precision with different cell sizes, we can see from the figure that the smaller size gives lower uncertainty.

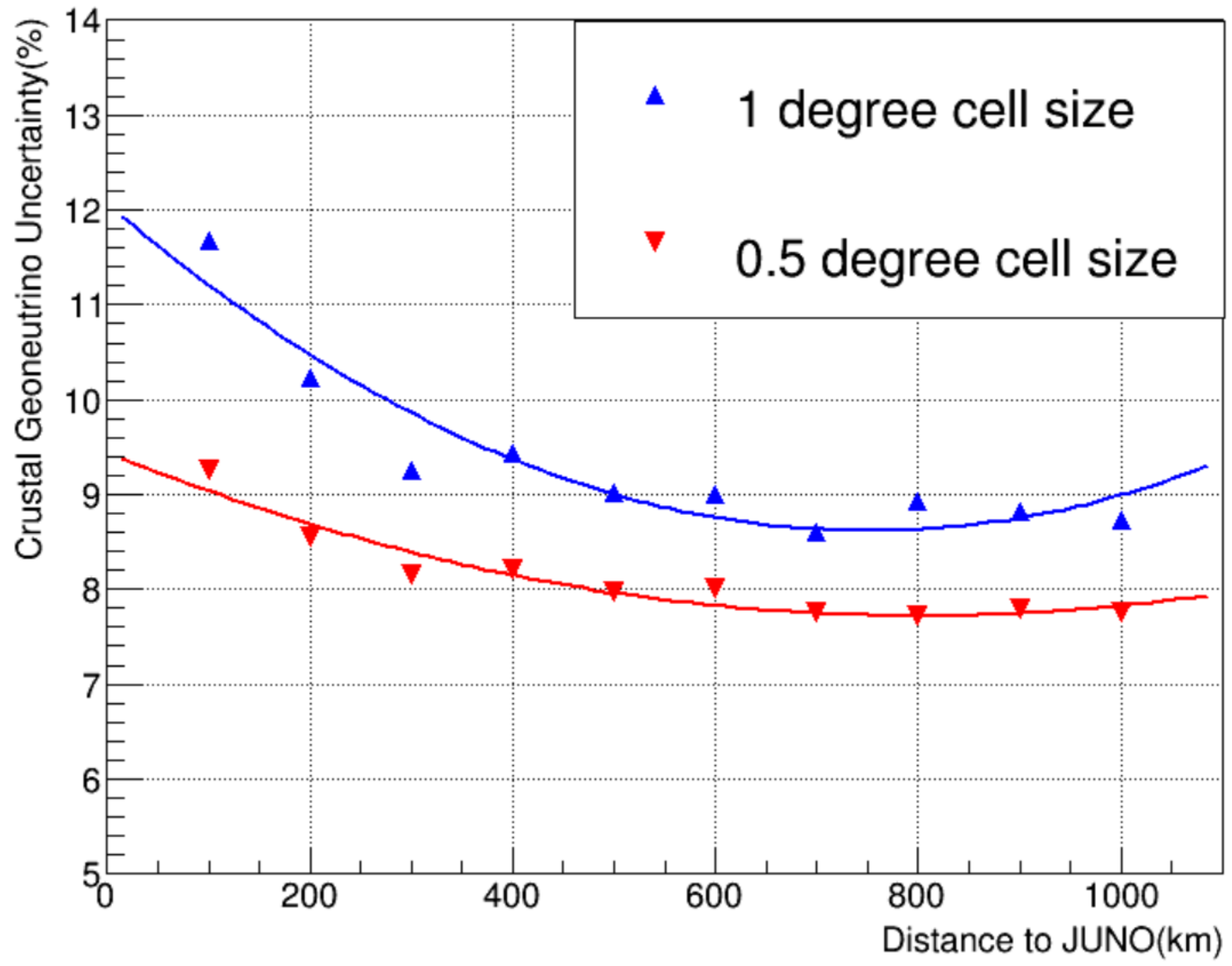

***Figure B2.*** *The predicted crustal geoneutrino uncertainties vs distance to JUNO. The blue upper triangle at 1°×1° cell size, the red down triangle at 0.5°×0.5° cell size.*

## Appendix C: Geoneutrino Signal, Mass of Uranium and Radiogenic Heat

This relationship between geoneutrino signal and masses of heat generating elements like uranium will be developed in the following way:

(1) After excluding the local 10° × 10° region, the proximity argument (Fiorentini et al., 2007 provides for the signal from the rest of the world:

$$S_{RW}(U) = (4.3 + 14.76 \times m(U)) \pm (2.32 + 2.61 \times \mathrm{m}(U))$$

where the signal is in TNU, the mass is in units of $10^{17}$ kg. In the BSE model, the central value of m(U) is $0.8 \times 10^{17}$kg, the interval within the $\pm$ sign corresponds to the full range of models which have been considered (Fiorentini et al. 2007).

(2) The regional contribution from the local 10 ° × 10 °, calculated in Section 5 is

$$S_{reg}(U) = 21.3 \pm 4.0$$

(3) Combining the regional contribution $S_{RW}(U)$ and rest of the world result $S_{reg}(U)$, then the uranium geoneutrino signal as a function of uranium mass in the Earth is given by:

$$S(U) = S_0(U) \pm \delta(U)$$

where $S_0(U) = 25.6 + 14.76 \times m(U)$ and $\delta(U) = \sqrt{(2.61 \times \mathrm{m}(U) + 2.32)^2 + 4.0^2}$

Note that different sources of uncertainties are combined by the means of root square.

From the signal of U, the Th contribution can be calculated based on an assumed chondritic Th/U ratio. Given the abundances of U and Th and mass of a given layer in the Earth, the radiogenic heat production can be calculated as below following Fiorentini et al. (2007):

$$H_R(U + Th) = 9.85 \times \mathrm{m}(U) + 2.67 \times \mathrm{m}(Th)$$